\documentclass[final,3p,times,twocolumn]{elsarticle}

\usepackage{blindtext, graphicx, amsmath, algorithm, algpseudocode, pifont, algcompatible, comment, layout, amsthm, amssymb}
\usepackage{enumitem}   
\usepackage{eso-pic}
\usepackage{booktabs}
\usepackage{float}

\usepackage[utf8]{inputenc}
\usepackage[english]{babel}
\usepackage{hyperref} 
\hypersetup{ colorlinks=true, linkcolor=black, filecolor=black, urlcolor=cyan, }
\usepackage{multirow}
\usepackage{circledtext}
\usepackage[T1]{fontenc}
\usepackage{etoolbox}
\usepackage{subcaption}

\circledtextset{charf=\Large}
\def\revised{\textcolor{black}}

\journal{ICT Express}

\begin{document}

\begin{frontmatter}

\title{Prediction of Permissioned Blockchain Performance\\for Resource Scaling Configurations}

\author{Seungwoo Jung\corref{cor1}}
\ead{swjung@os.korea.ac.kr}

\author{Yeonho Yoo\corref{cor1}}
\ead{yhyoo@os.korea.ac.kr}

\author{Gyeongsik Yang\corref{cor2}}
\ead{g\_yang@korea.ac.kr}

\author{Chuck Yoo\corref{cor2}}
\ead{chuckyoo@os.korea.ac.kr}

\address{Department of Computer Science and Engineering, Korea University, Seoul, Republic of Korea}

\cortext[cor1]{These authors contributed equally to this work}
\cortext[cor2]{Co-corresponding authors}

\begin{abstract}
Blockchain is increasingly offered as blockchain-as-a-service (BaaS) by cloud service providers. However, configuring BaaS appropriately for optimal performance and reliability resorts to try-and-error. A key challenge is that BaaS is often perceived as a ``black-box,'' leading to uncertainties in performance and resource provisioning. Previous studies attempted to address this challenge; however, the impacts of both vertical and horizontal scaling remain elusive. To this end, we present machine learning-based models to predict network reliability and throughput based on scaling configurations. In our evaluation, the models exhibit prediction errors of $\sim$1.9\%, which is highly accurate and can be applied in the real-world.

2018 The Korean Institute of Communications and Information Sciences. Publishing Services by Elsevier B.V. This is an open access article under the CC BY-NC-ND license (http://creativecommons.org/licenses/by-nc-nd/4.0/).
\end{abstract}


\end{frontmatter}

\section{Introduction}\label{sec:1}

Recently, blockchain technology has gained attention in both industry and academia due to its various advantages, such as data integrity, security, and auditability \cite{yang2022resource}. At present, public cloud providers such as Google Cloud \cite{googleBlockchainNode} and AWS \cite{amazonScalableBlockchain} have released managed blockchain services, called BaaS (blockchain-as-a-service), and numerous blockchain networks (or nodes) are deployed in public clouds using BaaS.

Utilizing BaaS to support a permissioned blockchain needs to consider the performance of the entire network. Specifically, it is still not clear how to configure the number of blockchain nodes (e.g., peers and orderers) that constitute a blockchain network (horizontal scaling) and the computing resources, such as CPU, for these nodes (vertical scaling). It is well known that these horizontal and vertical scaling factors significantly affect the performance and reliability of the blockchain network \cite{thakkar2021scaling}, such as the number of data transactions stored (committed) in the blockchain per second (transactions per second, TPS) and the number of failed transactions (transaction success rate). Thus, configuring BaaS is often treated as blackbox problem \cite{guggenberger2022depth}.

Several studies observed this problem and attempted to analyze the performance of BaaS. Yang et al. \cite{yang2022resource} analyzed the BaaS performance across various consensus protocols. Thakkar et al. \cite{thakkar2021scaling} benchmarked the performance of vertical and horizontal scaling on the scalability of Hyperledger Fabric (HLF). However, although these studies provided an analysis of BaaS configurations, they failed to predict the performance of new BaaS systems. This gap means that the scaling configuration process is still ambiguous and a blackbox problem. Another study by Sukhwani et al. \cite{sukhwani2018performance} used stochastic reward nets to theoretically model the performance of HLF v1.0+. However, their model's use in BaaS configuration is limited. As a theoretical model, modeling requires a deep understanding of the stochastic reward net and internal processes of HLF is required. Thus, the adaptability and flexibility of the model are poor, and the predictions on new and unseen blockchain networks are quite restricted.

To overcome the limitations of previous studies, we introduce new machine learning (ML)-based prediction models to estimate the performance of permissioned blockchains based on scaling factors. To the best of our knowledge, this is the first approach that leverages ML to consider scaling configurations in permissioned blockchains, thereby facilitating easier decision-making on scaling factors. Specifically, we devise two ML models to predict 1) transaction success rate and 2) throughput (TPS), which are the two primary metrics in planning and evaluating blockchain networks \cite{hyperledgerHyperledgerCaliper}. Our models incorporate scaling factors on horizontal and vertical scaling (e.g.,  number of peers and amount of CPU resources) as input features for prediction.

To train these models, a dataset that includes performance metrics, specifically transaction success rate and throughput, across various scaling factors, is required. However, to the best of our knowledge, such datasets are lacking. Therefore, we build our own dataset using a representative permissioned blockchain platform, HLF\cite{hyperledgerHyperledgerFabric}. By measuring the performance metrics under diverse scaling factors, we compile a dataset containing 593 records and train the two prediction models using this dataset.

Our evaluation of the models shows a prediction error of up to 1.9\%, which demonstrates their high accuracy and practicality for real world applications. In addition, we demonstrate the use-cases of these models with the scaling factors not used in model training. Specifically, we demonstrate two use-cases: 1) determining the optimal scaling configurations for a given throughput and 2) predicting the maximum throughput for a configured blockchain network. These two use-cases highlight the practicality of the models. 


\vspace{-1em}
\section{Background}\label{sec:2}


\begin{figure}[]
    \centering
    \begin{minipage}[valign=c]{.22\columnwidth}
        \centering
        \includegraphics[width=\columnwidth]{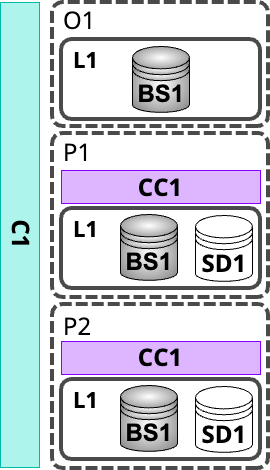}\vspace{-.5em}
        \caption{HLF components}
        \label{fig:hlf-component}
    \end{minipage}\hfill
    \begin{minipage}[valign=c]{.76\columnwidth}
        \centering
        \includegraphics[width=\columnwidth]{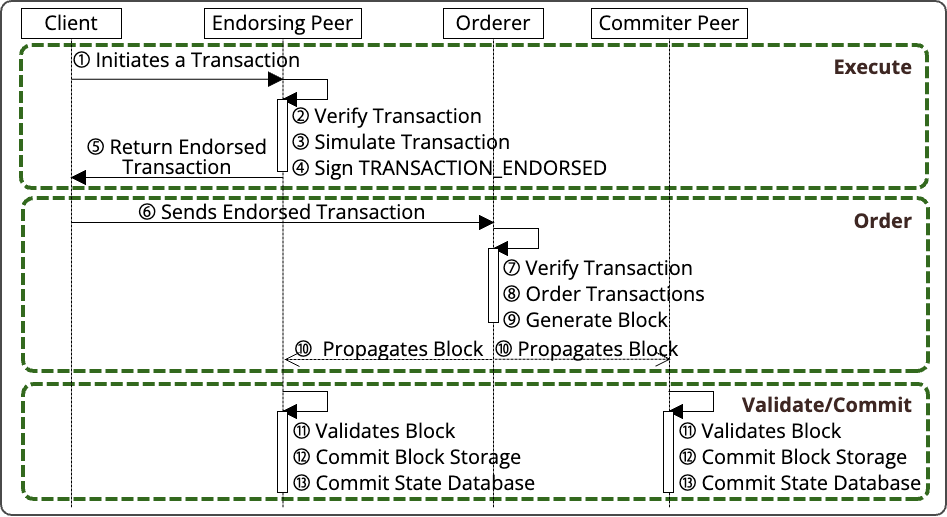}\vspace{-.5em}
        \caption{HLF transaction flow}
        \label{fig:hlf-flow}
    \end{minipage}\vspace{-1.5em}
\end{figure}


\subsection{HLF Components}\label{sec:2:1}

The key components of HLF are the channel and two types of nodes: peer and orderer. Fig. \ref{fig:hlf-component} shows a blockchain network with one channel (C1), two peer nodes (P1 and P2), and one orderer node (O1). We explain them in detail.
First, a channel represents a single blockchain network on the HLF. The peer nodes and orderer nodes participate in the channel. In Fig. \ref{fig:hlf-component}, P1, P2, and O1 participate in C1. Each channel has a ledger (L1) that is replicated by every orderer and peer.

Next, a peer is the endpoint at which clients (users) access the blockchain. Peers join more than one channel and perform the execution and endorsement (to be explained in detail at Section \ref{sec:2:2}) for the transaction requests submitted by clients. Each peer consists of block storage (BS1 in Fig. \ref{fig:hlf-component}), state DB (SD1), and chaincode (CC1). Block storage saves the group of transactions (blocks) of the channel. State DB stores the current state of the ledger, called the world state. Chaincode is a program that is executed when a new transaction request arrives. \revised{There are two types of peers: endorsing peers, which execute chaincode for client transaction requests, and committer peers, which verify transactions.}

In addition, the orderer receives transactions that have collected sufficient endorsements from peers, and it generates blocks by determining the order of the endorsed transactions through a consensus algorithm. Also, the orderer maintains a block storage to ensure the order and consistency of the created blocks.

\subsection{Transaction Flow of HLF}\label{sec:2:2}

Fig. \ref{fig:hlf-flow} shows the transaction flow of HLF that consists of three phases: execute, order, and validate. When clients request transactions, 1) peers execute (endorse) transactions, 2) orderers organize transactions with a consensus algorithm and generate blocks, and 3) peers validate blocks and update (commit) them into the ledger. We explain each phase in detail below.
 
\revised{Execute phase begins when the endorsing peer receives a transaction request (\circledtext{1} in Fig. \ref{fig:hlf-flow}). The peer verifies the transaction's signature (\circledtext{2}), simulates the chaincode (\circledtext{3}), and signs the transaction, creating an endorsed transaction (\circledtext{4}). It then returns the endorsed transaction to the client (\circledtext{5}).}

\revised{Order phase is the subsequent phase of the execute phase. Once a client has collected enough endorsed transactions, the client sends the transactions to the orderer (\circledtext{6}). The orderer receives the transactions and verifies whether the transactions contain a sufficient number of endorsements and whether the signatures are valid (\circledtext{7}). It then organizes the transactions in order using a consensus algorithm (\circledtext{8}) and generates a block (\circledtext{9}). The ordering phase ends by propagating the generated block to all other peers in the channel (\circledtext{10}).}

Validate/Commit phase is the last phase of transaction processing flow. When a peer receives the block from the orderer, both endorsing and committer peers perform several verifications on the blocks that check 1) whether the signatures of the transactions are correct, 2) whether the transactions contain sufficient endorsements, and 3) whether the data remains unchanged from the execute phase (\circledtext{11} in Fig. \ref{fig:hlf-flow}). After these verifications, the new block is added to block storage (\circledtext{12}), and the state DB is updated (\circledtext{13}).

\subsection{Scalability of HLF}\label{sec:2:3}

HLF operates the execute-order-validate phases for transactions. In contrast, other blockchain methods \cite{ethereumEthereumWhitepaper} first order transactions and then execute them so that they have to perform the same processes in both the order and execute phases. Compared to them, in HLF, not all nodes are required to undertake the execution process. Instead, they only need to obtain the required number of endorsements (results of the execute phase) from other peers, as required by the chaincode endorsement policy. Furthermore, endorsements in the execute phase can be processed in parallel. This distinction allows HLF to enhance its performance by distributing the peers, thus enabling horizontal scaling. For example, when the blockchain network requires only one endorsement to execute a transaction, the execution requests are distributed as 1/n, where n peers exist.

\revised{
However, HLF does not exhibit linear scalability in the platform performance, such as throughput or transaction success rate, when the number of peers increases (horizontal scaling). We observed the non-linear scalability in our experiments, but we had to omit the detailed results due to the space limit. Another study \cite{thakkar2021scaling} also reported the non-linear scalability with the increasing number of peers where the throughput did not increase linearly. We find that this scalability is akin to the vertical scaling as well. Because the scalability is not linear, it is difficult and important to predict the scaling configurations, such as the number of peers and CPU quotas, which this paper addresses.}

\vspace{-1em}
\section{Data Collection and Model Training}\label{sec:3}

\subsection{Data Collection Setup}\label{sec:3:1}

\begin{table}[]
    \begin{minipage}[t]{.48\linewidth}
\caption{Host resources.}\label{table1}\vspace{-1em}
\resizebox{\columnwidth}{!}{%
\begin{tabular}{l||l}
\hline
\textbf{CPU}     & \begin{tabular}[c]{@{}l@{}}Intel Xeon E5-2650 v4 \\@ 2.20 GHz (24 cores)\end{tabular}              \\ \hline
\textbf{RAM}     & 64 GB                                                    \\ \hline
\textbf{Network} & 10 GbE                                                 \\ \hline
\textbf{OS}      & Ubuntu 18.04.6 LTS \\ \hline
\end{tabular}%
}
    \end{minipage}%
    \hfill
    \begin{minipage}[t]{.51\linewidth}
\caption{HLF setup.}\label{table2}\vspace{-1em}
\resizebox{\columnwidth}{!}{%
\begin{tabular}{l||l}
\hline
\textbf{Configuration} & \textbf{Value} \\ \hline
Endorsement policy                                                            & 1-of-N                         \\ \hline
Block size                                                                     & 10                             \\ \hline
\# of channels and orderers & 1                              \\ \hline
CPU quota of orderer                                                              & 2                              \\ \hline
\end{tabular}%
}
    \end{minipage} \vspace{-1.5em}
\end{table}

\textbf{Machine setup.} \revised{
We use two separate machines with identical specifications as Table \ref{table1}: one is for running HLF that configures a blockchain network, and the other is for running Hyperledger Caliper, a de-facto benchmark to measure the performance of HLF. The reason for running HLF components on a dedicated machine is to isolate the blockchain performance from external factors such as the network. Thus, we can measure the performance changes only due to its own scaling factors. Moreover, running Hyperledger Caliper consumes significant computing resources that easily interfere with the performance of HLF. Thus, this configuration isolates the performance from external factors.}

\textbf{HLF setup.} We use a 1-of-N endorsement policy, which means that each transaction is endorsed by only one peer. This policy allows for rapid distribution of transaction requests as the number of peers increases. So, this policy clearly shows the performance gains through horizontal scaling. In addition, we set the block size (number of transactions per block), the number of channels, number of orderers, and CPU quota of orderer to 10, 1, 1, and 2, respectively. These values are determined by empirical experiments to ensure that the configurations do not introduce any bottlenecks in performance measurements or scaling factor changes. We generate 20000 transactions every time, each performing bubble sort. This is a common operation in evaluating HLF \cite{saingre2020bctmark}. To ensure a clean blockchain state, we always reset the entire blockchain network for each experiment. Table \ref{table2} summarizes the HLF setup.

\subsection{Data Collection}\label{sec:3:2}

We measure each data record that consists of output features (labels) and input features. For the output features, we evaluate two metrics that represent the performance of HLF as follows: 
\begin{itemize}[noitemsep,topsep=0pt]
    \setlength\itemsep{0em}
    \item \textbf{Success rate:} percentage of transactions that are successfully processed (committed) out of the total number of requested transactions.
    \item \textbf{Throughput:} number of transactions committed (validated) by the blockchain network per second. 
\end{itemize}

\revised{For the input features, we choose representative vertical and horizontal scaling factors of HLF \cite{thakkar2021scaling, guggenberger2022depth} because the scaling factors affect the success rate and throughput. For vertical scaling, we choose CPU quota per peer, and for horizontal scaling, we select the number of peers. We also include the transaction request rate. The reason is that the transaction request rate indicates the speed of transactions entering the blockchain network per second, so it determines success rate and throughput. For example, when the number of peers and CPU quota are both set to 1, we observe that the success rate fluctuates by ~89.2\% when the transaction request rate is increased from 250 to 550. This fluctuation is also similarly observed in the throughput.}

\revised{We considered the other factors, such as endorsement policy and block size, but excluded them because other studies \cite{thakkar2021scaling, guggenberger2022depth} also identified CPU quota per peer, number of peers, and transaction request rate (three features of our model) as the factors that change the success rate and throughput for scaling.}

\revised{For dataset collection, we vary the three input features within the following ranges. First, the CPU quota per peer varies by the number of physical cores. Our machine has 24 cores, so when we run five peers, we change the CPU quota per peer from one to four as five cores per peer exceed the capacity of the machine. Second, the number of peers is increased from 1 to 10 in order to ensure the allocation of at least two CPU quotas for each peer.}

\revised{Lastly, we change the transaction request rates of 250, 350, 450, and 550. We set this range to avoid network overhead and to capture the scaling effect clearly. In total, we collect 593 data records by changing the three input features and measuring the two output features.}

\subsection{Model Training}\label{sec:3:3}

\begin{figure}[]
    \centering
    \begin{minipage}[c]{.62\columnwidth}
        \centering
        \caption{Hyperparameter boundaries and selected values.}\label{table:model_parameters}\vspace{-.5em}
        \resizebox{\columnwidth}{!}{%
        \begin{tabular}{@{}c|c|c|c@{}}
        \hline\hline
            \textbf{Model} & \textbf{Hyperparameter} & \begin{tabular}[c]{@{}c@{}}\textbf{Search}\\ \textbf{range}\end{tabular} & \begin{tabular}[c]{@{}c@{}}\textbf{Selected}\\\textbf{value}\end{tabular} \\
            \hline\hline
            \multirow{4}{*}{\begin{tabular}[c]{@{}c@{}}Success rate \\ model\end{tabular}} & n\_estimators & 1--50 & 5 \\
            & max\_depth & 1--30 & 5 \\
            & min\_samples\_split & 1--100 & 4 \\
            & min\_samples\_leaf & 1--50 & 2 \\
            \hline
            \multirow{4}{*}{\begin{tabular}[c]{@{}c@{}}Throughput \\ model\end{tabular}} & n\_estimators & 1--50 & 19 \\
            & max\_depth & 1--30 & 16 \\
            & min\_samples\_split & 1--100 & 2 \\
            & min\_samples\_leaf & 1--50 & 2 \\
            \hline\hline
        \end{tabular}%
        }
    \end{minipage}\hfill
    \begin{minipage}[c]{.37\columnwidth}
        \centering
        \includegraphics[width=\columnwidth]{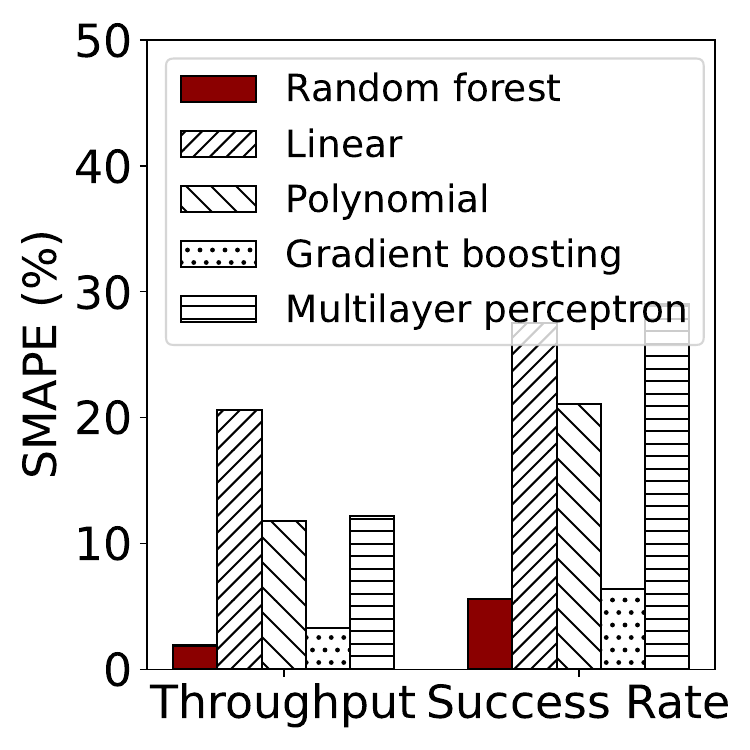}\vspace{-1em}
        \caption{Prediction accuracy comparison}
        \label{fig:model_acc}\vspace{-.5em}
    \end{minipage}\vspace{-1.5em}
\end{figure}

We train two random forest models: one predicts the success rate, and the other predicts throughput. For training, we split our data records into two: 80\% for training models and the remaining 20\% for evaluating the model accuracy. We test different types of ML models that are widely applied in performance prediction \cite{hameed2021machine, lee2007methods}, such as multiple linear regression and polynomial regression models. In this study, we use the random forest model as it shows the highest accuracy, which will be explained in Section \ref{sec:4:1}.

\revised{Random forest models have several hyperparameters that make differences in prediction accuracy, such as the number of trees (n\_estimators), the maximum depth of the trees (max\_depth), the minimum number of samples required to split an internal node (min\_samples\_split), and the minimum number of samples required to be at a leaf node (min\_samples\_leaf). We determine the hyperparameters by grid search \cite{yang2020hyperparameter} that tests all possible values within the specified boundaries and selects the best value for prediction accuracy. We choose a grid search because we observe that, for the random forest model, even a slight change in hyperparameters impacts the prediction accuracy. For example, in the success rate prediction model, when n\_depth is set to 1, test set accuracy shows 61.6\%. However, when n\_depth is set to 5, test set accuracy shows 97.2\%. Thus, through grid search, we test all possible configurations to achieve better accuracy. Table \ref{table:model_parameters} lists the types of hyperparameters, their search boundaries, and the selected best values for the prediction models’ accuracy.}


\vspace{-1em}
\section{Performance of Prediction Models}\label{sec:4}

Here, we evaluate the trained prediction models. 
The models are trained with sklearn v1.3.0 using the same machine that is used for data collection (Section \ref{sec:3:1}). 


\subsection{Model Accuracy}\label{sec:4:1}


\revised{We evaluate the accuracy of the two types of models by training them using ML algorithms: linear regression, polynomial regression, gradient boosting regression, multi-layer perceptron regression, and random forest regression. The prediction accuracy is calculated using symmetric mean absolute percentage error (SMAPE) \cite{smape}. A SMAPE value of 10\% means that, on average, the predicted values differ from the ground truth by 10\%. Lower SMAPE indicates better model accuracy, with 0\% indicating the perfect accuracy. Fig. \ref{fig:model_acc} shows the prediction errors of the five algorithms. Among them, random forest regression shows the lowest errors for success rate and throughput prediction at 5.6\% and 1.9\%, respectively, which are 3.8$\times$ and 6.38$\times$ better than the other algorithms on average.}

\subsection{Model Use-cases}\label{sec:4:2}

Next, we present use-cases that demonstrate how the two models mitigate the black box issues in scaling blockchain networks as BaaS in real-world clouds.


\textbf{Case 1: finding optimal scaling configurations.} When attempting to find the optimal scaling configuration to meet desired performance requirements of a blockchain network, the current approach in most cloud environments is to use repetitive trial-and-error processes. This is due to the absence of tools or methods to predict performance before provisioning and running all possible cases of the scaling configurations.\vspace{-.5em}
\begin{equation}
    \begin{gathered}
    (n_c, n_p) = \arg \min_{n_c, n_p} (n_c \times w_c + n_p \times w_p)\\
    where \quad n_c \in N_c \quad and \quad n_p \in N_p
    \label{math:modeling:grid_search}
    \end{gathered}
\end{equation}\vspace{-1em}

We describe the trial-and-error process as Equation (1). There are six notations: $n_c$ and $n_p$ represent the CPU quota and the peer values, each. $N_c$ and $N_p$ represent the sets of all possible values for $n_c$ and $n_p$. Also, $w_c$ and $w_p$ represent the weights given to the CPU quota and number of peers. These weights reflect the cost of CPU quota (vertical scaling) or peers (horizontal scaling). For example, when adding one vCPU costs \$1 and adding one VM costs \$2, $w_c$ is given as 1 and $w_p$ as 2. Thus, Equation (1) represents the process of 1) testing all possible $n_c$ and $n_p$ values with their weights ($w_c$ and $w_p$) and 2) selecting the pair whose cost is the minimum.

We assume that our blockchain network on the cloud receives 500 requests per second on average, indicating that its transaction request rate is 500. The performance requirements of the system are as follows: a success rate of 100\% and a throughput higher than 90\% of the transaction request rate ($500 \times 0.9 = 450$). We use the two prediction models to determine the proper configurations for the system.

\begin{figure}
\centering
\begin{minipage}[valign=t]{.66\columnwidth}
\centering
\captionsetup[subfigure]{justification=centering}
  \begin{subfigure}[]{0.49\textwidth}
    \includegraphics[width=\textwidth]{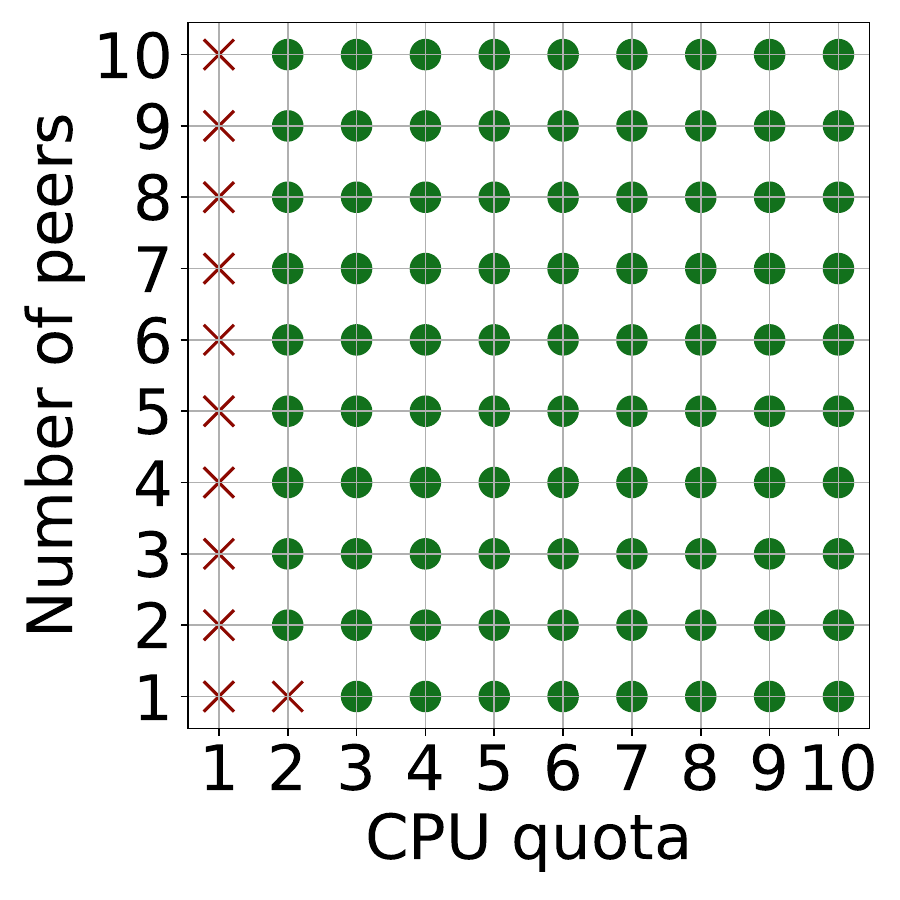}
    \caption{Success rate}\label{fig:modeling:predicted:succ_rate}
  \end{subfigure}
  \begin{subfigure}[]{0.49\textwidth}
    \includegraphics[width=\textwidth]{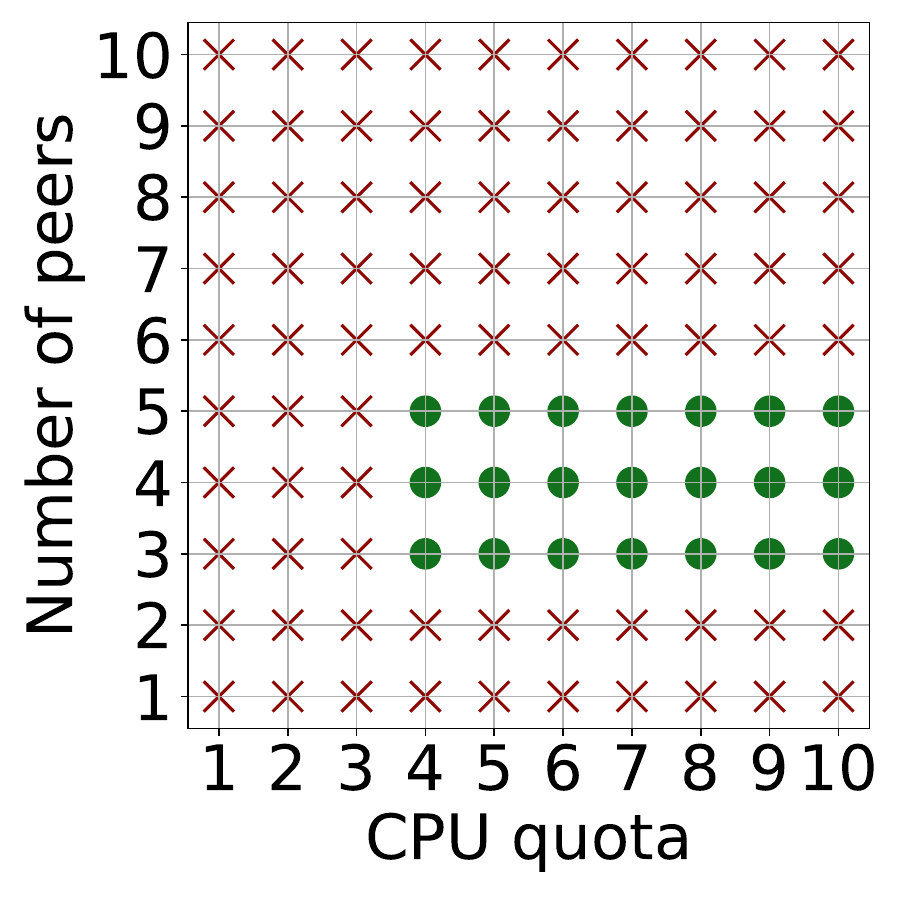}
    \caption{Throughput}\label{fig:modeling:predicted:throughput}
  \end{subfigure}\vspace{-1em}
  \caption{Predicted performances (500 transaction request rate).}\label{fig:modeling:predicted}
\end{minipage}\hfil
\begin{minipage}[valign=t]{.33\columnwidth}
\includegraphics[width=\columnwidth]{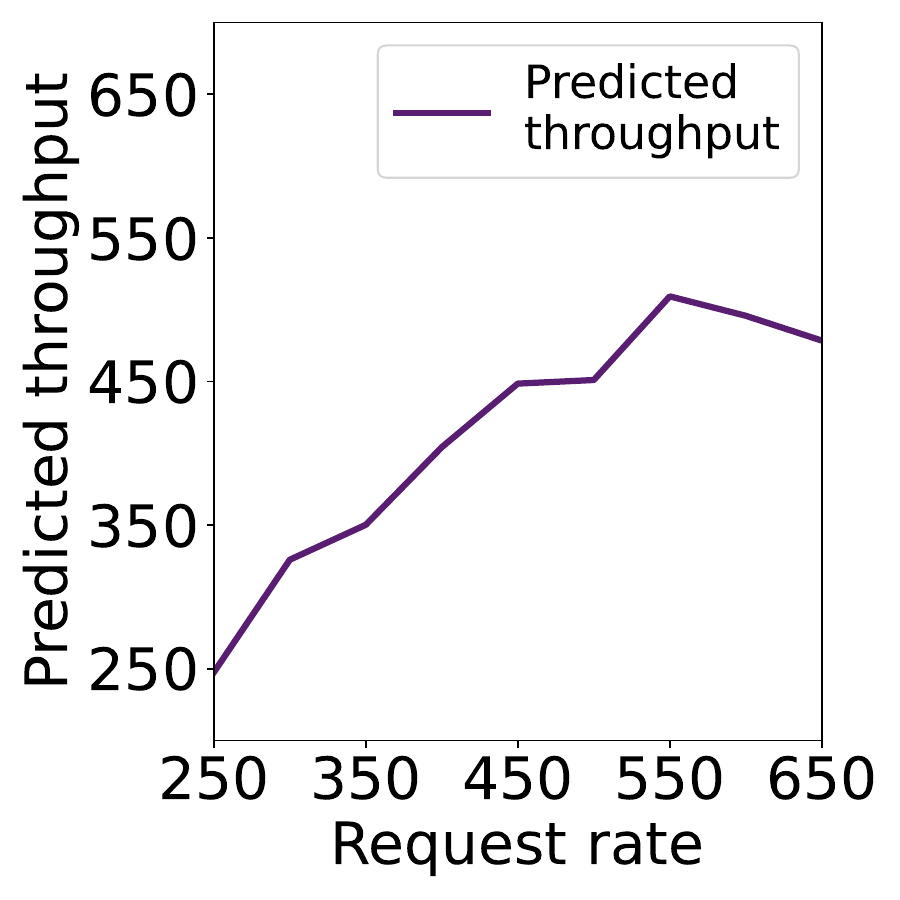}\vspace{-1em}
\caption{Predicted throughput (4 CPU quotas, 4 peers).}
\label{fig:modeling:predicted_max}
\end{minipage}\vspace{-1.5em}
\end{figure}

Figs. \ref{fig:modeling:predicted:succ_rate} and \ref{fig:modeling:predicted:throughput} present the predictions for the success rate and throughput, respectively. The x-axis and y-axis of the figures represent the CPU quota and number of peers, respectively. We mark each point with two symbols: $\times$ symbol when the predicted performance does not satisfy the desired performance requirement, and $\bigcirc$ when satisfied. For example, in Fig. \ref{fig:modeling:predicted:succ_rate}, the point where the x-axis is 1 and the y-axis is 1 is marked with $\times$, indicating a success rate of less than 100\%. In Fig. \ref{fig:modeling:predicted:throughput}, the point where the x-axis is 5 and y-axis is 5 is marked with $\bigcirc$, meaning that the throughput is higher than 450.

Using the prediction results, we can consider only the points where $\bigcirc$ appears as possible scaling configurations without any real provisioning or execution of BaaS. 
If $w_c$ and $w_p$ are the same (e.g., 1), we can select three peers and four CPU quotas that satisfy both the success rate and throughput requirements. \revised{We also test our example using a real HLF setup, achieving a 100\% success rate and a transaction rate of 495.3 TPS, confirming that all requirements are met.}

\textbf{Case 2: throughput predictions.} For Baas engineers, it is important to predict the performance of the deployed blockchain network. This prediction enables the engineers to make informed decisions in advance about whether to scale their blockchain based on their expected transaction demands. As a means of predicting performance, we utilize the two trained models. We assume that we have deployed a BaaS network, scaled with both the CPU quotas and peers set to four. In this system, we aim to determine how the throughput changes when the upcoming transaction request rates vary from 250 to 650.
 
Fig. \ref{fig:modeling:predicted_max} shows the predicted throughput values when the transaction request rates vary. The throughput increases up to 509.3 when the transaction request rate increases to 550. However, beyond this rate, the throughput begins to decrease until the transaction request rate reaches 650.
We also validate the predicted performance by measuring the actual performance of a real system. We measure four representative request rates---300, 400, 500, and 600---that are not used for model training. The measured throughput values are 298.9, 398.6, 497.2, and 497.9, respectively. By comparing the measured performance with the predicted ones in Fig. \ref{fig:modeling:predicted_max}, the average prediction error is 5\%, which is reasonable.

Based on the accurate prediction results, BaaS engineers can predict that, in a configuration with four peers and four CPU quotas, the maximum throughput is 509.3 when the transaction request rate is 550. If the BaaS system is expected to receive a transaction request rate that is higher than 550, scale-in for the system is required.

\vspace{-1em}
\section{Related Work}\label{sec:5}

Existing studies mainly focused on performance and resource consumption analysis and revealed several limitations. Thakkar et al. \cite{thakkar2018performance} provided an analysis of HLF v1.1, and another work \cite{thakkar2021scaling} identified scalability bottlenecks and suggested a validation scheme on a subset of transactions for scalability. Despite the efforts, previous studies lacked performance prediction methods for HLF. Sukhwani et al. \cite{sukhwani2018performance} modeled performance metrics; however, success rates were not covered, and it has practicality issues due to the complexity of using stochastic reward nets. In contrast, our study introduces ML-based models for horizontal and vertical scaling, with a recent platform and practical use-cases.

\vspace{-1em}
\section{Discussion}\label{sec:discussion}

\revised{\textbf{Model overfitting.} We use K-fold cross-validation to check the overfitting of the trained models, which provides the baseline prediction accuracy when overfitting is prevented \cite{nti2021performance}. So, if the accuracy from K-fold cross-validation is similar to the accuracy of our trained models, it indicates that our models are far from overfitting. K-fold cross-validation splits the dataset into K equal parts (folds) and uses a different fold for validation each time. Thus, even using the same dataset, the model is trained and validated K times, each time using a different fold for validation (to determine model convergence). We choose K=10 as \cite{malakouti2023usage}. The result is 7.4\% SMAPE on average for the success rate model and 1.4\% for the throughput model. These values differ by only 1.2\% from our model training results in Section \ref{sec:4:1}, thus showing that our models are not overfitting.}

\revised{\textbf{Prediction model on public blockchain.} Our work focuses on the scaling factors of private blockchains. In public blockchain platforms, the number of peers or the resources per peer cannot be determined, as public blockchains allow free participation of any device. Therefore, our prediction models are designed for private blockchains, where the number of peers (horizontal scaling) or the CPU quota (vertical scaling) can be configured. For future work, we plan to design the performance prediction models for public blockchains by considering their configuration parameters \cite{consensus, blocktime}, such as consensus mechanisms and block time.}

\revised{\textbf{Different platforms.} Among private blockchain platforms, we choose HLF as the base platform to develop prediction models because it is the de-facto platform for running BaaS on clouds, such as IBM Cloud \cite{ibmBlockchainEnterprise}, AWS \cite{amazonScalableBlockchain}, and GCP \cite{googleBlockchainNode}. Other studies on blockchain performance \cite{yang2022resource, thakkar2021scaling, sukhwani2018performance} have used HLF. We believe that our prediction models can be trained and validated for different platforms, such as Hyperledger Besu and Ethereum, because they handle transactions similar to HLF, and the scaling factors in transaction processing are identical \cite{abdella2021architecture, fan2020performance}. For future work, we plan to extend our models to other platforms.}

\vspace{-1.5em}
\section{Conclusion}\label{sec:6}

This study presents two ML-based performance prediction models for transaction success rate and throughput. Our prediction models exhibit prediction errors of ~1.9\% for various horizontal and vertical scaling factors, which is highly accurate. Furthermore, we present two useful cases in which prediction models can benefit BaaS users by enabling them to determine scaling configurations or predict their performance. 

\vspace{-1.5em}
\section*{Acknowledgments}

This work was supported by Basic Science Research Program through National Research Foundation of Korea (NRF) funded by Ministry of Education (NRF-2021R1A6A1A13044830), by NRF grant funded by Korea government (MSIT) (NRF-2023R1A2C3004145, RS-2024-00336564), by ICT Creative Consilience Program through IITP (RS-2020-II201819), and by Google Cloud Research Credits program. We used generative AI only to improve the readability of this paper in part.


\bibliographystyle{elsarticle-num}
\vspace{-0.3cm}

\bibliography{yks-bib}
\end{document}